\documentclass[aps,twocolumn,superscriptaddress,floatfix,tightenlines,notitlepage]{revtex4-2}

\usepackage{graphicx}
\usepackage{dcolumn}
\usepackage{bm}
\usepackage[nice]{nicefrac}

\usepackage{amsmath, amssymb}    
\usepackage[breaklinks=true,hidelinks]{hyperref}

\newcommand{\icts}{International Centre for  Theoretical Sciences, Tata Institute of Fundamental Research,  Bangalore 560089, India}
\newcommand{\iitk}{Department of Mechanical Engineering, Indian Institute of Technology Kanpur, Kanpur 208016, India}

\begin{document}

\title{Hydrodynamics of Dense Active Fluids: Turbulence-Like States and the Role of Advected Activity}

\author{Sandip Sahoo}
 \email{sandipsahoo09902@gmail.com}
 \affiliation{\icts}
 
\author{Siddhartha Mukherjee}
 \email{smukherjee@iitk.ac.in}
 \affiliation{\iitk}
 
\author{Samriddhi Sankar Ray}%
\email{samriddhisankarray@gmail.com}
\affiliation{\icts}

\begin{abstract}

	Dense suspensions of self-propelled bacteria and related active fluids
	exhibit spontaneous flow generation, vortex formation, and
	spatiotemporally chaotic dynamics despite operating at vanishingly
	small Reynolds numbers. These phenomena, commonly referred to as active
	turbulence, display striking visual and statistical similarities to
	classical inertial turbulence while arising from fundamentally
	different nonequilibrium mechanisms. In this article, we present a
	combined review and theoretical study of hydrodynamic models for dense
	active fluids, with particular emphasis on bacterial suspensions
	described by the Toner--Tu--Swift--Hohenberg (TTSH) framework. We
	review key experimental and theoretical developments underlying the
	analogy between active and inertial turbulence, highlighting the
	emergence of multiple dynamical regimes and the conditions under which
	universal spectral and intermittent behavior arises in homogeneous
	systems. Moving beyond the conventional assumption of spatially uniform
	activity, we introduce a minimal model in which the activity field is
	heterogeneous and dynamically advected by the flow it generates.
	Thus treating activity as a spatiotemporally evolving field coupled to the
	TTSH dynamics, we investigate how advection and diffusion lead to sharp
	activity fronts, confinement of turbulent motion, and complex
	interfacial morphologies. Our numerical results demonstrate that
	spatial variations in activity can induce transient coexistence of
	distinct spectral regimes and that universality in active turbulence is
	inherently local and time-dependent in heterogeneous systems. These
	findings underscore the importance of treating activity as a dynamical
	field in its own right and provide a framework for studying active
	turbulence in more realistic, spatially structured biological and
	synthetic active matter systems.

\end{abstract}


\maketitle

\section{Introduction}

Active matter systems consist of agents that convert stored or ambient energy
into systematic motion at the microscopic
scale~\cite{toner2005hydrodynamics,Ramaswamy2010,Vicsek2012,marchetti2013hydrodynamics,ramaswamy2017active,alert2022active}.
Prominent examples include bacterial suspensions~\cite{wu2000particle},
cytoskeletal extracts driven by molecular motors~\cite{Sanchez}, synthetic
Janus particles~\cite{Howse}, and vibrated granular media~\cite{Narayan2007}.
Unlike equilibrium systems, active matter is intrinsically out of equilibrium:
Detailed balance is broken locally, and energy injection occurs continuously at
the level of individual constituents. As a result, active systems exhibit
collective phenomena --- such as flocking~\cite{Vicsek,Choudhary2015,Gupta},
swarming~\cite{ZhangSwarm,Peruani2012,ariel2015swarming}, phase
separation~\cite{Tailleur,Fily,Redner2013}, and spontaneous flow
generation~\cite{Simha,Voituriez2005,Marenduzzo,Jain} --- that have no direct
equilibrium analogues.

Among these systems, dense suspensions of swimming bacteria occupy a special
place. When bacterial concentration is sufficiently high, individual swimming
trajectories become strongly coupled through steric interactions, near-field
hydrodynamics, and self-generated flow
fields~\cite{marchetti2013hydrodynamics,alert2022active}.  Experiments on
quasi-two-dimensional bacterial films have repeatedly shown the emergence of
vortices, jets, and spatiotemporally chaotic patterns at length scales much
larger than the size of a single swimmer. Remarkably, these flows arise in a
regime where the Reynolds number, based on both swimmer size and collective
velocity scales, remains exceedingly small. This apparent contradiction ---
turbulence-like motion in the absence of inertia and at negligible Reynolds
numbers--- has motivated extensive experimental, numerical, and theoretical
investigation over the past
decade~\cite{alert2022active,wu2000particle,ariel2015swarming,wensink2012meso,dunkel2013fluid,bratanov2015new,james2018turbulence,mukherjee2021anomalous,singh2022lagrangian,mukherjee2023intermittency,kiran2024onset}.

Early theoretical descriptions of collective motion in active systems focused
on flocking and long-range orientational order, most notably within the
Toner--Tu framework~\cite{TonerTu95,TonerTu98}. These theories successfully capture the onset of
collective motion and the emergence of polar order in dry active systems, where
momentum is not conserved. However, dense bacterial suspensions typically do
not settle into uniformly ordered states. Instead, they exhibit sustained
chaotic dynamics characterized by the continuous creation and annihilation of
vortices, a well-defined mesoscale length, and broadband velocity fluctuations.
Capturing these features requires continuum descriptions that go beyond
homogeneous flocking and explicitly incorporate pattern-forming instabilities~\cite{marchetti2013hydrodynamics,alert2022active}.

A significant advance in this direction was the realization that active
stresses generated by extensile swimmers can effectively act as negative
viscosity, destabilizing long-wavelength modes of the flow. When supplemented
by higher-order stabilizing gradients, this leads naturally to a hydrodynamic
description closely related to the Swift--Hohenberg
equation~\cite{SwiftHohenberg,CrossHohenberg}, but augmented by the nonlinear
advection and symmetry-breaking terms characteristic of active matter. The
resulting Toner--Tu--Swift--Hohenberg (TTSH)
framework~\cite{wensink2012meso,dunkel2013fluid}, which we review in the next
section, has emerged as a minimal continuum model capable of reproducing many
qualitative and quantitative features of dense bacterial turbulence, including
vortex
formation~\cite{bratanov2015new,james2018turbulence,mukherjee2021anomalous,singh2022lagrangian,mukherjee2023intermittency,kiran2024onset}.

Parallel to these theoretical developments, an analogy between active
turbulence and classical inertial turbulence gradually emerged. Early
experimental and numerical studies focused primarily on energy spectra in dense
bacterial flows and reported apparent power-law scaling over intermediate
ranges~\cite{wensink2012meso,dunkel2013fluid,bratanov2015new}. However, the corresponding scaling exponents were found to depend
sensitively on the level of activity, and the flows were largely
non-intermittent, indicating only a partial and qualitative resemblance to high
Reynolds number turbulence.

Subsequent investigations~\cite{mukherjee2021anomalous,singh2022lagrangian,mukherjee2023intermittency,kiran2024onset} refined this picture by revealing a more nuanced
scenario. In particular, it was shown that beyond a critical level of activity,
active turbulent flows undergo a qualitative transition: The scaling exponents
of the energy spectra become universal~\cite{mukherjee2023intermittency} and essentially independent of activity,
while strong intermittency develops in both Eulerian and Lagrangian statistics~\cite{mukherjee2023intermittency,kiran2024onset}.
In this high-activity regime, active turbulence exhibits statistical features
that more closely parallel those of inertial turbulence, despite the absence of
inertia and the fundamentally different mechanisms of energy injection and
dissipation. Active turbulence thus spans multiple dynamical regimes and
constitutes a distinct, yet deeply connected, class of chaotic flow phenomena~\cite{alert2022active}.

Most existing theoretical and numerical studies of active turbulence, however,
rely on a crucial simplifying assumption: The activity is taken to be spatially
uniform and temporally constant~\cite{alert2022active}. This assumption is
reasonable for well-mixed bacterial suspensions under controlled laboratory
conditions, but it is rarely satisfied in natural or biologically relevant
settings~\cite{Mukherjee2025}. In real bacterial colonies, activity can vary due to nutrient
gradients, oxygen availability, or local
crowding~\cite{thar2003bacteria,arlt2018painting,frangipane2018dynamic,arlt2019dynamics,nishiguchi2018engineering,reinken2020organizing,reinken2022ising}.
Similarly, in synthetic active systems, spatial modulation of activity can be
externally imposed through light, chemical cues, or patterned
substrates~\cite{palacci2013living,schuppler2016boundaries,ross2019controlling,zhang2021spatiotemporal,zhang2021autonomous,shankar2022topological,lemma2023spatio,hoff2009prokaryotic,wilde2017light}.
These considerations raise fundamental questions: How does spatially
heterogeneous activity modify the structure and statistics of active
turbulence? Can activity itself behave as a dynamically evolving field,
transported by the flow it generates? And, to what extent can such an activity
field be viewed as an active or passive scalar in a turbulent-like environment?

In this work, we combine a semi-review of active
turbulence with new theoretical and numerical results on heterogeneously active
suspensions. We first recall the pedagogy of the TTSH hydrodynamic
framework for dense bacterial suspensions and review key developments
underpinning the analogy between active and inertial turbulence. We then move
beyond the constant-activity paradigm and introduce models in which the
activity field is spatially varying and dynamically advected by the flow. We investigate how such advected activity fields evolve, homogenize, or form sharp fronts, and we
characterize the resulting flow structures and scaling properties.

The paper is organized as follows. In Sec.\eqref{sec:TTSH}, we derive the
Toner--Tu--Swift--Hohenberg hydrodynamic equations appropriate for dense
bacterial suspensions. Section~\eqref{sec:analogy} reviews the development of the turbulence
analogy in active matter, focusing on key experimental and theoretical
milestones. In Sec.\eqref{sec:Heterogeneous_activity}, we discuss the limitations of uniform-activity models
and motivate the need for heterogeneous descriptions. Section~\eqref{sec:advected_activity} introduces a
minimal advection model for spatially varying activity. Our main results are
presented in Sec.\eqref{sec:Results}, where we analyze the evolution, mixing, and front dynamics
of the activity field and their impact on active turbulence. Finally, Sec.\eqref{sec:Conclusions}
summarizes our findings and outlines open questions and future directions.

\section{Hydrodynamic Theory for Dense Bacterial Suspensions: A Short Review}
\label{sec:TTSH}

Dense suspensions of swimming bacteria exhibit collective dynamics that differ
qualitatively from both dilute active gases and equilibrium fluids. At
sufficiently high volume fractions, steric interactions, near-field
hydrodynamics, and self-generated flows lead to the emergence of mesoscale
vortices, jets, and spatiotemporally chaotic patterns commonly referred to as
\emph{bacterial turbulence}. Despite the low Reynolds number of the individual
swimmers, these systems display features reminiscent of inertial turbulence,
albeit governed by entirely different physical mechanisms.

A successful continuum description of such systems must therefore (i)
incorporate spontaneous polar ordering due to self-propulsion, (ii) account for
momentum injection at the swimmer scale, and (iii) include nonlinear saturation
and pattern-selection mechanisms. These requirements are naturally met by a
hybrid hydrodynamic framework combining elements of the Toner--Tu theory of
flocking with Swift--Hohenberg-type gradient terms. In this section, we briefly review
this Toner--Tu--Swift--Hohenberg (TTSH) model starting from symmetry arguments
and coarse-grained conservation laws~\cite{wensink2012meso,dunkel2013fluid}; we refer 
the reader to the review article by Alert, Casademunt, and Joanny~\cite{alert2022active} for 
a detailed modern review of this field.

We consider a dense suspension of self-propelled bacteria confined to a thin film or quasi-two-dimensional geometry, as is common in experiments. The relevant coarse-grained fields are:
\begin{itemize}
	\item the bacterial number density $\rho(\mathbf{r},t)$,
	\item the polarization (or mean velocity) field $\mathbf{v}(\mathbf{r},t)$.
\end{itemize}

At high densities, density fluctuations are strongly suppressed due to steric constraints, allowing one to work in the incompressible or weakly compressible limit. We therefore assume $\rho \simeq \rho_0$ constant and focus on the dynamics of $\mathbf{v}$, which simultaneously represents the local orientational order and the coarse-grained flow field. Number conservation implies
\begin{equation}
	\partial_t \rho + \nabla \cdot (\rho \mathbf{v}) = 0,
\end{equation}
which reduces to $\nabla \cdot \mathbf{v} = 0$ in the incompressible limit.

In the absence of momentum conservation (neglecting friction with the substrate or a surrounding fluid), the velocity field obeys relaxational rather than Navier--Stokes dynamics. The most general equation of motion for $\mathbf{v}$, consistent with rotational invariance and broken Galilean invariance, is given by the Toner--Tu equation:

\begin{equation}
\begin{aligned}
	\partial_t \mathbf{v}
	+ \lambda_1 (\mathbf{v}\cdot\nabla)\mathbf{v}
	&+ \lambda_2 (\nabla\cdot\mathbf{v})\mathbf{v}
	+ \lambda_3 \nabla |\mathbf{v}|^2 \\
	&= -\nabla p
	- (\alpha + \beta |\mathbf{v}|^2) \mathbf{v}
	+ D \nabla^2 \mathbf{v}.
	\label{eq:TT}
\end{aligned}
\end{equation}

Here, $\alpha < 0$ controls the linear instability toward collective motion,
while $\beta>0$ saturates the growth of $\mathbf{v}$. The coefficients
$\lambda_i$ encode the absence of Galilean invariance typical of active matter
systems~\cite{wensink2012meso}. The pressure $p$ enforces incompressibility.
For $\alpha<0$, Eq.~\eqref{eq:TT} admits a homogeneous polar state
$|\mathbf{v}| = \sqrt{|\alpha|/\beta}$, corresponding to coherent flocking.
However, dense bacterial suspensions rarely settle into such uniform states.
Instead, they exhibit vortex patterns with a well-defined characteristic length
scale, which the Toner--Tu theory alone cannot capture.

Experimental studies indicate that active stresses generated by extensile swimmers effectively reduce the viscosity of the suspension. At high activity, this can lead to a \emph{negative effective viscosity}, destabilizing long-wavelength modes. To account for this, the diffusion term is generalized as $\Gamma_0 \nabla^2 \mathbf{v}$,
with $\Gamma_0 < 0$. A linear stability analysis of Eq.~\eqref{eq:TT} around $\mathbf{v}=0$ then yields a growth rate
\begin{equation}
	\sigma(k) = -(\alpha + \Gamma_0 k^2),
\end{equation}
which is \emph{positive} for all wavelengths as long as $\alpha < 0$ and $\Gamma_0 < 0$. This implies unbounded growth of modes, and hence ill-posed dynamics unless higher-order stabilizing terms are included.

To regularize the instability and select a finite vortex scale, we introduce the leading higher-order gradient term $\Gamma_2 \nabla^4 \mathbf{v}$ allowed by symmetry with $\Gamma_2 > 0$.
The resulting linear dispersion relation becomes
\begin{equation}
	\sigma(k) = -(\alpha + \Gamma_0 k^2 + \Gamma_2 k^4),
\end{equation}
which is maximized at a finite wavenumber
\begin{equation}
	k_c = \sqrt{\frac{|\Gamma_0|}{2\Gamma_2}}.
\end{equation}

This is the hallmark of Swift--Hohenberg dynamics and leads to spontaneous pattern formation with a characteristic length scale $\ell_c \sim k_c^{-1}$, consistent with experimental observations of bacterial vortices.

Collecting all terms, the full TTSH equation for dense bacterial suspensions reads
\begin{equation}
	\partial_t \mathbf{v}
	+ \lambda_1 (\mathbf{v}\cdot\nabla)\mathbf{v}
	= -\nabla p
	- (\alpha + \beta |\mathbf{v}|^2) \mathbf{v} + \Gamma_0 \nabla^2 \mathbf{v}
	- \Gamma_2 \nabla^4 \mathbf{v},
	\label{eq:TTSH}
\end{equation}
supplemented by the incompressibility constraint $\nabla\cdot\mathbf{v}=0$.

Equation~\eqref{eq:TTSH} constitutes a minimal continuum model for mesoscale
bacterial turbulence. The nonlinear advection term transfers energy across
scales, while the competition between $\Gamma_0<0$ and $\Gamma_2>0$ injects
energy at a finite wavelength. The cubic nonlinearity saturates the instability
and prevents blow-up.

The TTSH framework provides a unified description of dense active fluids that
bridges polar flocking and pattern-forming instabilities. It is valid on lengthscales that are large compared to the swimmer sizes and timescales longer than their
reorientation time. While derived phenomenologically, several coefficients can
be related to microscopic parameters such as swim speed, dipolar force
strength, and rotational diffusivity.

Importantly, the model leads to an inverse energy flux in two dimensions and
reproduces key statistical features of bacterial turbulence, including
non-Gaussian velocity distributions and scale-dependent energy spectra. As
such, it has emerged as a canonical hydrodynamic description for dense
bacterial suspensions and related active fluids.

\section{Analogy Between Active and Inertial Turbulence: Key Developments}
\label{sec:analogy}

A central theme in active fluid research is the analogy between
spatiotemporally chaotic flows in dense suspensions and classical, high
Reynolds number turbulence. Although the underlying physical mechanisms differ
--- active flows are driven by internal energy injection at the microscale,
whereas inertial turbulence emerges from nonlinear momentum transfer from large to small scales at high
Reynolds numbers --- several statistical and structural similarities have
motivated comparative studies. Below we review a few key contributions that cement
and critically evaluate this analogy.

In a landmark study~\cite{wensink2012meso}, Wensink and collaborators combined experiments on dense
\emph{Bacillus subtilis} suspensions with particle simulations and continuum
theory to characterize what they termed ``mesoscale turbulence in living fluids''. They measured energy spectra and velocity structure functions
in quasi-two-dimensional geometries and found broad-band kinetic energy across
intermediate wavenumbers, reminiscent of inertial turbulence spectra. Their
experiments revealed vortical structures with a characteristic mesoscopic scale
and nontrivial correlations in the flow field, suggesting that chaotic
bacterial dynamics could share statistical signatures with conventional
turbulence despite the absence of inertia. Based on these observations, they
proposed a minimal continuum model for incompressible active bacterial flow and
demonstrated that it reproduces many experimentally observed statistical
features of active turbulence.

This study laid the groundwork for viewing bacterial turbulence not simply as
random motion but as a distinct nonequilibrium phase with well-defined
statistical properties. Importantly, it catalyzed subsequent theoretical
efforts to generalize active hydrodynamics beyond simple Toner-Tu models to
incorporate turbulence-like spatiotemporal
chaos~\cite{james2018turbulence,mukherjee2021anomalous,singh2022lagrangian,mukherjee2023intermittency,kiran2024onset,james2018vortex,james2021emergence,Kiran2023,Jain2025,Kashyap2025}.

Bratanov \emph{et al.}~\cite{bratanov2015new} formulated a more systematic comparison between active
turbulent flows and inertial turbulence by extending continuum descriptions to
include both the usual Navier-Stokes nonlinearity and active stress
contributions allowed by symmetry. They demonstrated that the interaction
between the quadratic convective term from Navier-Stokes and the cubic
nonlinearity stemming from active stresses leads to energy spectra at large
scales exhibiting power-law regimes, albeit with nonuniversal exponents
dependent on system parameters and finite-size effects. 

This result highlighted that although active turbulence shares some spectral
features with inertial turbulence, such as cascades or broad power-law-like
ranges, the scaling laws and universality classes can differ fundamentally. In
particular, active turbulence defines a distinct class of turbulent flows
where energy injection occurs at the microscale via self-propulsion, and the
scaling exponents are set by non-equilibrium organization rather than inertial
transfer alone. 

Building on these insights, Mukherjee~\textit{et al.}~\cite{mukherjee2023intermittency} and 
Kiran~\textit{et al.}~\cite{kiran2024onset} investigated the emergence
of a universal chaotic regime in active turbulence when activity exceeds a
critical level. Their studies suggest that, beyond a threshold of activity,
active flows transition into highly intermittent states with non-Gaussian
velocity distributions~\cite{kiran2024onset} and chaotic fluctuations that mimic classical turbulence
intermittency --- albeit driven by internal energy injection rather than
inertial mechanisms. Such work indicates that active turbulence may exhibit its
own version of intermittency universality, not tied directly to classical
Kolmogorov scaling and phenomenology, but rather emergent from the interplay between active
drive, nonlinear advection, and dissipation. 

Using such hydrodynamic models, several studies have also shown that Lagrangian
statistics~\cite{mukherjee2021anomalous,singh2022lagrangian,kiran2024onset,Pandit2025}
in active turbulence bear unique signatures of the underlying nonequilibrium
dynamics. For example, analysis of Lagrangian tracer trajectories reveals
super-diffusive scaling and L\'evy walk-like behavior over intermediate time
scales, in stark contrast to the near-diffusive behavior typical of high
Reynolds number flows~\cite{mukherjee2021anomalous,singh2022lagrangian}. This super-diffusion can obscure the crossover from
ballistic to diffusive motion and is tied to transient coherent structures in
the flow that actively mold the tracer dynamics. Such features underscore key
distinctions between active and inertial turbulent transport, even when
Eulerian velocity fields exhibit similar-looking chaotic patterns. 

Collectively, such studies illustrate both the power and the
limits of the analogy between active and inertial turbulence. On the one hand,
dense bacterial suspensions can exhibit spectra, mesoscale vortical organization, chaotic transport and universal turbulence statistics at sufficiently high activity, reminiscent of high Reynolds number flows. On the other hand,
key features such as nonuniversal scaling, L\'evy-type Lagrangian statistics,
and activity-induced intermittency highlight fundamental differences rooted in
the mechanisms of energy injection and dissipation. The active turbulence
paradigm hence stands as a distinct universality class of chaotic fluid motion,
connected to but not subsumed by classical turbulence theory.

\section{Beyond a constant-activity paradigm: Heterogeneous suspensions}
\label{sec:Heterogeneous_activity}

Most theoretical and numerical studies of active turbulence assume that the
activity parameter is spatially uniform and constant in time. This simplifying
assumption has played a crucial role in establishing minimal hydrodynamic
descriptions and identifying key dynamical regimes, as illustrated in the previous section. However, this paradigm represents an idealization that is rarely realized in experimental or biological settings.

In practice, activity in active matter systems is almost always
heterogeneous~\cite{thar2003bacteria,arlt2018painting,frangipane2018dynamic,arlt2019dynamics,nishiguchi2018engineering,reinken2020organizing,reinken2022ising}.
In bacterial suspensions, local activity depends on nutrient availability,
oxygen concentration, waste accumulation, and local crowding, all of which can
vary significantly in space and time. Similarly, in synthetic active systems,
activity can be externally modulated using light, chemical gradients, or
patterned substrates, leading to controlled but intrinsically heterogeneous
activity landscapes. Even in nominally uniform experimental preparations,
finite-size effects and boundary interactions inevitably generate spatial
variations in activity.

Despite its experimental relevance, the impact of spatially varying activity on
active turbulence has only recently begun to be explored in a systematic
manner. A notable step in this direction is the recent work of Mukherjee
\emph{et al.}~\cite{Mukherjee2025}, which explicitly relaxes the
assumption of homogeneous activity and investigates dense active flows subject
to spatially heterogeneous driving. Motivated by experimental situations in
which activity is localized or externally patterned~\cite{yang2019quenching,arlt2019dynamics,zhang2021spatiotemporal,hokmabad2025spatial,martinez2025interfacial}, this study demonstrates
that activity heterogeneity fundamentally alters the structure of active
turbulent states. In particular, regions of high activity act as localized
sources of chaotic motion, while surrounding weakly active regions remain
comparatively quiescent, leading to the emergence of sharp dynamical interfaces
separating distinct flow regimes.

An important insight from this work is that these interfaces are not static
boundaries but fluctuate strongly in time, 

The most important insight from this work is that these interfaces exhibit rich dynamics, exhibiting 
intermittency and a complex spatio-temporal structure. As a result, heterogeneous activity introduces new physical ingredients into active turbulence, including confinement of
vorticity, interface-dominated transport, and nontrivial mixing between active
and inactive regions. Such effects have no direct analogue in homogeneous
models and cannot be captured by treating activity as a single global control
parameter.

At the same time, the activity heterogeneity considered in
Ref.~\cite{Mukherjee2025} is quenched in space, representing
systems in which activity variations are imposed externally and remain fixed in
time. While this is a natural and experimentally relevant starting point, particularly with the aim of active matter control, many
active systems operate in a different regime, where activity itself can be
transported, deformed, and mixed by the flow it generates. In bacterial
suspensions, for instance, local activity can spread through advection of
bacteria, while diffusion and growth processes concurrently act to smoothen or
sharpen activity gradients.

This observation motivates a complementary perspective in which activity is
treated as a dynamical field rather than a prescribed spatial pattern. In such
a setting, the flow not only responds to activity gradients but also actively
reshapes them, leading to a two-way coupling between turbulence and activity
distribution. Understanding how this coupling modifies flow organization,
intermittency, and spectral properties requires going beyond both
homogeneous-activity models and quenched-heterogeneity approaches.

In the following sections, we adopt this viewpoint and study active turbulence
in the presence of an advected, spatially varying activity field. This
framework allows us to bridge homogeneous active turbulence and systems with
fixed activity heterogeneity, and to address how transient fronts, mixing, and
homogenization of activity influence the emergence, persistence, and eventual
loss of turbulent behavior.

\section{An advection model for heterogeneously active turbulence}
\label{sec:advected_activity}

We introduce a minimal continuum framework in which the
activity field is treated as a spatiotemporally evolving scalar coupled to the
Toner--Tu--Swift--Hohenberg dynamics, which makes activity not
merely a static control parameter but a dynamical quantity that can be
transported, deformed, and mixed by the flow it generates. 
We consider the activity parameter $\alpha$ appearing in Eq.~\eqref{eq:TTSH} to
be a coarse-grained field $\alpha(\mathbf{x},t)$, representing the local
strength of active driving. Physically, this field may encode variations in
swimmer concentration, metabolic activity, nutrient availability, or externally
imposed control parameters such as light intensity in synthetic active systems.
In contrast to homogeneous models, where $\alpha$ is taken to be constant, here
$\alpha(\mathbf{x},t)$ is allowed to vary continuously in space and time. In
the absence of flow, spatial variations in activity would relax through
diffusive processes associated with microscopic motion, growth, or chemical
transport. In the presence of a chaotic velocity field, however, activity
gradients can be strongly distorted and advected, leading to complex
spatiotemporal structures analogous to those encountered in scalar mixing
problems.

To capture these effects in the simplest possible manner, we model the evolution of the activity field 
using an advection--diffusion equation,
\begin{equation}
	\partial_t \alpha + \mathbf{v}\cdot\nabla \alpha
	= \kappa \nabla^2 \alpha ,
	\label{eq:alpha_advection}
\end{equation}
where $\mathbf{v}(\mathbf{x},t)$ is the velocity field governed by
Eq.~\eqref{eq:TTSH}, and $\kappa$ is an effective diffusivity associated with
activity transport. Equation~\eqref{eq:alpha_advection} represents a minimal description
in which activity is transported by the flow while undergoing diffusive
smoothing at small scales. No explicit source or sink terms are included,
allowing us to isolate the role of advection and diffusion in shaping activity
heterogeneity and to focus on how turbulent-like flows redistribute activity.

The relative importance of advective stretching and diffusive regularization is conveniently characterized by Schmidt number, $\displaystyle \mathrm{Sc} = \frac{|\Gamma_0|}{\kappa}$, where $\Gamma_0$ denotes a characteristic viscosity scale of the velocity
field. Large Schmidt numbers correspond to weak diffusion, in which activity
gradients can persist and sharpen under the action of the flow, while small
Schmidt numbers lead to smoother and more rapidly homogenized activity fields.

The coupling between the activity field and the velocity dynamics arises
entirely through the dependence of Eq.~\eqref{eq:TTSH} on the local value of
$\alpha(\mathbf{x},t)$. Spatial variations in activity therefore modulate the
linear instability and energy injection into the flow, leading to regions of
the domain that support qualitatively different dynamical regimes. In turn, the
resulting velocity field advects, stretches, and mixes the activity field
through Eq.~\eqref{eq:alpha_advection}. This two-way coupling establishes a
nontrivial feedback loop: Activity heterogeneity organizes the flow by
localizing or suppressing turbulent motion, while the flow continuously
reshapes the activity distribution.

In this sense, the activity field behaves as an \emph{active scalar}. Although
its evolution equation resembles that of a passively advected field, its
spatial distribution directly controls where and how turbulent dynamics are
generated. This minimal framework allows us to systematically investigate how
the interplay between advection, diffusion, and activity-dependent forcing
gives rise to evolving activity fronts, flow confinement, and modifications of
spectral and statistical properties. 

\begin{figure*}[t]
	\includegraphics[width=\linewidth]{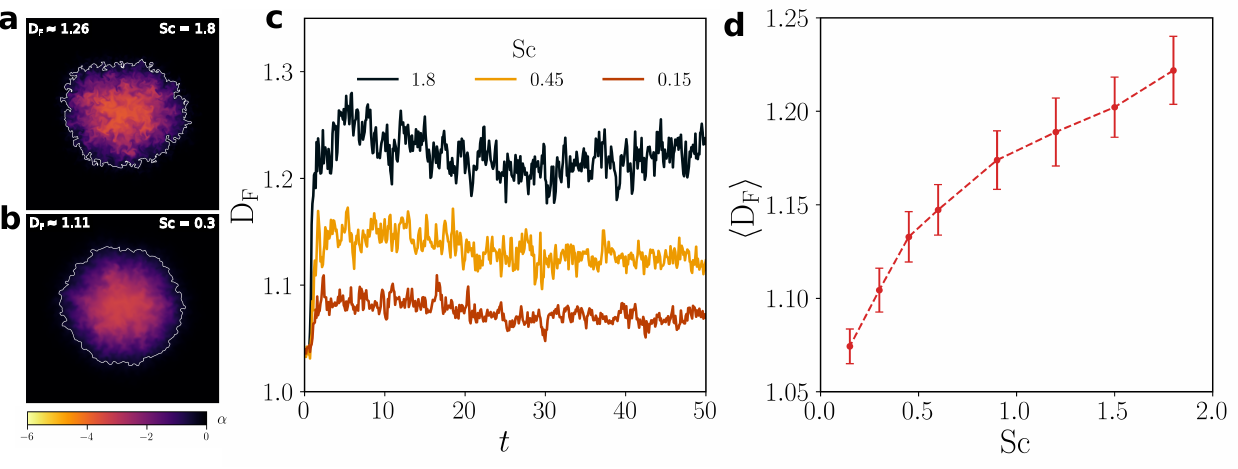}
	\caption{Representative plots of the activity field $\alpha$ at time $t=15$ for
		Schmidt numbers (a) $\mathrm{Sc}=1.8$ and (b) $\mathrm{Sc}=0.3$, with
		the white curve denoting an iso-contour of the activity field defining
		the evolving activity front. The Schmidt number controls the morphology
		of this front, which becomes increasingly convoluted as diffusion
		weakens. This is best seen in a time evolution of the scalar field for a large range of Schmidt numbers as shown 
		in Ref.~\cite{scalar_contours}. Panel (c) shows the temporal evolution of the fractal
		dimension $D_F$ of the activity front for different Schmidt numbers,
		illustrating a rapid growth from the initially smooth interface
		followed by saturation to a statistically steady value.  Panel (d)
		shows the corresponding steady-state values of $\langle D_F \rangle$,
		with error bars characterizing statistical fluctuations, as a function
		of the Schmidt number. The activity fronts become more fractal with
		increasing $\mathrm{Sc}$.}
	\label{Fig:Contours} 
\end{figure*}

We perform direct numerical simulations (DNSs) of the coupled equations \eqref{eq:TTSH} and \eqref{eq:alpha_advection} on square periodic domains of length $L = 20$ and $30$, discretized over $N^2 = 1024^2$ and $2048^2$ collocation points, respectively. Although the results presented here are drawn from the larger simulation, they are consistent across both system sizes. A standard  $1/2$-de-aliased pseudo-spectral algorithm is applied to \eqref{eq:TTSH} to account for the cubic non-linearity in the coarse-grained velocity field, while a $2/3$-de-aliased pseudo-spectral algorithm handles the quadratic nonlinearity arising from the activity advection by the velocity field in \eqref{eq:alpha_advection}. The time integration is carried out using a second-order Integrating Factor Runge-Kutta (IFRK2) scheme with a time step $\delta t = 0.0002$. The simulation parameters are kept identical to those used in the previous studies: $\Gamma_0 = - 0.045$, $\Gamma_2 = |\Gamma_0|^3$, $\beta =  0.5$ and $\lambda_1 = 3.5$. The heterogeneous activity field is initialized with $\alpha_{\mathrm{in}}=-6$ within a central disc of radius $r=5$ and $\alpha_{\mathrm{out}}= 0$ in the outside region. This configuration is particularly useful for analyzing the front morphologies and flow organization. A second initialization --- with $\alpha_{\mathrm{in}}=-8$ inside a disc of radius $r=10$ and $\alpha_{\mathrm{out}}= 4$ outside --- is employed to investigate the spectral features of the flow while preserving the underlying physics.

\section{Results}
\label{sec:Results}

\begin{figure*}
	\includegraphics[width=\linewidth]{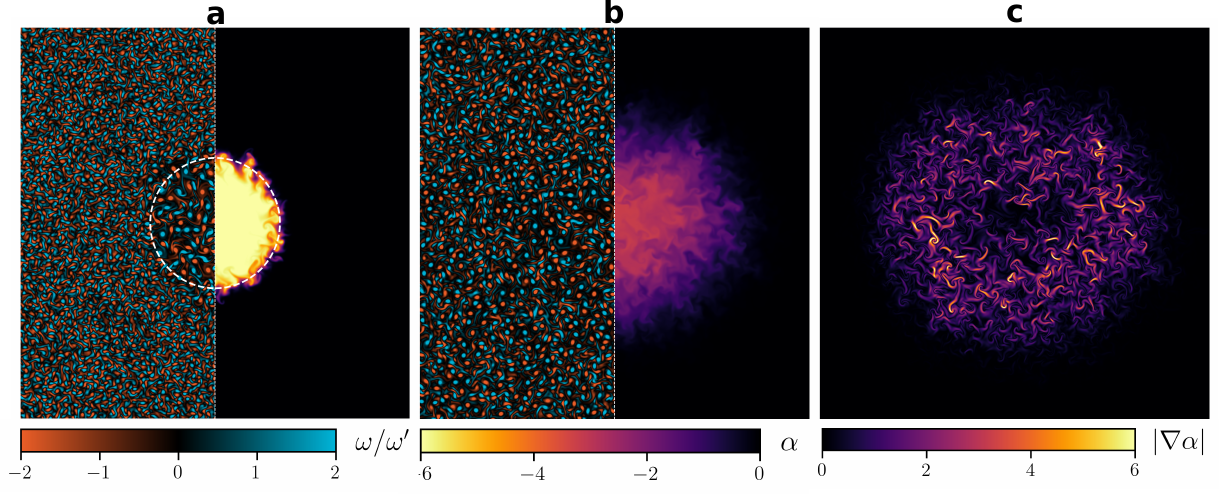}
	\caption{Representative plots of the (normalized) vorticity field (left panels)
		and activity field (right panels) for Schmidt number $\mathrm{Sc}=1.8$
		at (a) $t=1.0$ and (b) $t=20.0$. At early times, the turbulent flow is
		largely confined within the region of high activity, indicated by the
		white broken line in panel (a), while the surrounding region remains
		comparatively quiescent. At later times, as the activity spreads and
		becomes more homogeneous, this confinement weakens and vorticity
		structures extend more uniformly across the domain. Panel (c) shows the
		magnitude of the activity gradient field, $|\nabla \alpha|$, at $t=20$,
		revealing intermittent, worm-like structures associated with sharp
		activity fronts. We refer the reader to Ref.~\cite{fields} for a complete time evolution of the vorticity, activity and gradient of activity fields.}
	\label{Fig:Fields}
\end{figure*}

We now present results from numerical simulations of the
Toner--Tu--Swift--Hohenberg equation~\eqref{eq:TTSH}, coupled to a dynamically
evolving activity field~\eqref{eq:alpha_advection}. 
In contrast to the conventional homogeneous-activity
formulation, the activity $\alpha(\mathbf{x},t)$ here obeys an
advection--diffusion equation with diffusivity $\kappa$ and is transported by
the velocity field generated through Eq.~\eqref{eq:TTSH}.
The resulting dynamics are governed by a nontrivial feedback loop: Spatial
variations in activity modify the flow field, while the flow in turn advects,
stretches, and deforms the activity distribution. This interplay leads to the
formation of sharp activity fronts, heterogeneous flow structures, and complex
interfacial morphologies, as illustrated in Figs.~\ref{Fig:Contours} and~\ref{Fig:Fields}.

\subsection{Growth and morphological evolution of activity fronts}

We begin with a qualitative characterization of the activity field and its
associated fronts, focusing on the role of advection and diffusion in shaping
their evolution. The system is initialized with activity concentrated within a
circular region at the center of the domain, surrounded by a background of
lower activity. This configuration defines a well-posed initial interface
separating regions of high and low activity, allowing us to track the
subsequent growth and deformation of the activity front.

Representative snapshots of the activity field $\alpha(\mathbf{x},t)$ at a
fixed time are shown in Fig.~\ref{Fig:Contours}(a) and (b) for two different Schmidt numbers. For a complete 
visualisation of how such fields evolve for a wide range of Schmidt numbers, we refer the reader to the 
animation found in Ref.~\cite{scalar_contours}. At early
times, the initially smooth circular interface is rapidly distorted by the flow
generated within the active region. The activity patch is stretched, folded,
and advected outward, in close analogy with scalar mixing in turbulent flows.
We recall that similar interfaces are seen in high Reynolds number flows such 
as the boundaries between jets in $\beta$-plane zonal turbulence~\cite{sahoozonal} and turbulent/non-turbulent interfaces in mixing layers~\cite{bisset2002turbulent,westerweel2009momentum,chauhan2014turbulent,borrell2016properties,mazzino2021unraveling}.
As a result, the interface separating high and low-activity regions evolves
into a highly convoluted front, characterized by elongated filaments and sharp
gradients. 

The degree of interfacial roughening is strongly controlled by the Schmidt
number. For lower $\mathrm{Sc}$, diffusion acts efficiently to smoothen
gradients, leading to comparatively regular and rounded fronts. As
$\mathrm{Sc}$ is increased and diffusion weakens, advective stretching
dominates, producing increasingly intricate front geometries with fine-scale
structures. These fronts develop narrow filaments and deep indentations,
indicative of repeated stretching and folding by the underlying chaotic flow.

The growth of sharp activity gradients is evident from the activity front defined as the iso-contour $\alpha = - 0.4$, shown as white curves in Fig.~\ref{Fig:Contours}(a) and (b). These fronts
behave as dynamically evolving interfaces whose geometry reflects the
competition between advective deformation and diffusive smoothing. Importantly,
because activity directly feeds back onto the velocity field, the front is not
merely passively transported: Regions of strong gradients and high curvature
actively influence the local flow, further enhancing interfacial complexity.

To quantify this morphological evolution, we analyze the geometry of the
activity front using a box-counting approach to extract its fractal dimension.
The temporal evolution of the fractal dimension $D_F$ is shown in Fig.~\ref{Fig:Contours}(c) for
different Schmidt numbers. Starting from an initially smooth interface, $D_F$
grows rapidly as the front becomes increasingly wrinkled and irregular. After a
transient growth phase, the fractal dimension saturates to a statistically
steady value, indicating the emergence of distinct a self-similar interfacial morphology
during the mixing process.

Figure~\ref{Fig:Contours}(d) summarizes the dependence of the steady-state fractal dimension on
the Schmidt number. As expected, the fronts become more fractal as
$\mathrm{Sc}$ increases: Reduced diffusion allows the flow to sustain finer
interfacial structures before they are smoothed out. The systematic increase of
$D_F$ with $\mathrm{Sc}$ demonstrates that the geometry of activity fronts is a
robust outcome of the coupled advection--activity dynamics.

\subsection{Flow organization and confinement by heterogeneous activity}

\begin{figure*} \includegraphics[width=1.0\linewidth]{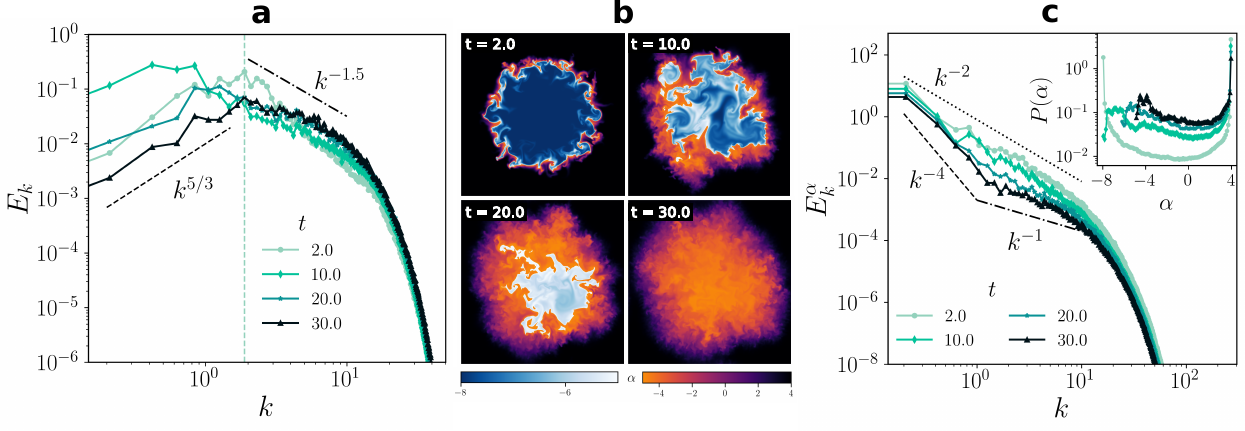}
	\caption{(a) Kinetic energy spectra $E(k)$ at four representative times
		for a system initialized with a strongly active patch of activity
		$\alpha_{\mathrm{in}}=-8$ inside a central disc of radius $r=10$ and
		$\alpha_{\mathrm{out}}=4$ outside. This extreme value is chosen to
		clearly expose the mechanism underlying dual scaling. At early times,
		the spectrum exhibits two distinct regimes: a universal $k^{-3/2}$
		scaling at larger wavenumbers, characteristic of strongly active
		turbulence ($\alpha<\alpha_c\approx -5$), and a non-universal scaling
		at smaller wavenumbers associated with weakly active regions. The
		crossover between these regimes occurs at a time-dependent wavenumber
		$k_R(t)\sim R^{-1}(t)$, where $R(t)$ is the effective linear size of
		the region with $\alpha<\alpha_c$. As shown in panel~(b), $R(t)$
		initially grows due to advection, causing $k_R(t)$ to shift to smaller
		values, and later decreases as diffusion and mixing erode the strongly
		active core, shifting $k_R(t)$ back to larger values and eventually
		eliminating the $k^{-3/2}$ regime.  (b) Snapshots of the activity field
		$\alpha(\mathbf{x},t)$ at the same four times, with regions satisfying
		$\alpha<\alpha_c$ highlighted in blue. These blue regions define $R(t)$
		and thus directly set the crossover scale $k_R(t)$ observed in
		panel~(a): their growth supports the coexistence of dual spectral
		scaling, while their eventual shrinkage explains the disappearance of
		the universal regime.  (c) The $\alpha$-spectrum shown at times as in panels (a) and (b) show distinct 
		scaling regimes $k^{-2}$ (dotted line), $k^{-4}$ (dashed line) and $k^{-1}$ (dot-dashed line). 
		(Inset) Probability density functions (PDFs) of
		$\alpha$ at these times, showing an initially near-bimodal
		distribution reflecting the coexistence of strongly and weakly active
		regions, which progressively flattens and eventually disappears for $\alpha \lesssim \alpha_c$,
		as the activity field becomes well mixed.}
	\label{Fig:Spectra}
\end{figure*}

While Fig.~\ref{Fig:Contours} emphasizes the evolution of activity fronts and their interfacial
morphology, Fig.~2 illustrates how the velocity and vorticity fields respond
to, and co-evolve with, the spreading activity. Together, these figures
highlight the two-way coupling between activity heterogeneity and flow
organization that distinguishes this system from homogeneous active turbulence.

Figures~\ref{Fig:Fields}(a) and (b) shows representative snapshots of the normalized vorticity field
(left panels) alongside the activity field (right panels) for a fixed Schmidt
number, $\mathrm{Sc}=1.8$, at two different times. At early times (Fig.~\ref{Fig:Fields}(a)),
the activity remains strongly localized within a ring-like region surrounding
the initial active patch. Correspondingly, the vorticity field is largely
confined to this high-activity domain, while the flow outside remains
comparatively quiescent. This reflects the role of activity as a local source
of instability and energy injection in the TTSH dynamics.

Within the active region, the vorticity field exhibits characteristic features
of highly active turbulence, including coherent vortices and elongated
streak-like structures.  In contrast, regions of lower activity support weaker,
less organized vorticity with distinctly different morphology. The sharp
transition between these two dynamical regimes closely follows the activity
front, emphasizing that the activity field effectively partitions the domain
into dynamically distinct subregions.

As time progresses and the activity front advances outward, this confinement
gradually weakens. Figure~\ref{Fig:Fields}(b) shows that at later times, once the activity has
spread over a larger fraction of the domain, the distinction between the flow
inside and outside the initially active region becomes less pronounced.
Vorticity structures extend more uniformly across the system, and the flow
approaches the statistically homogeneous active turbulence state associated with
the spatially averaged level of activity.

This gradual loss of confinement reflects the continuous reorganization of the
flow in response to the evolving activity distribution. As regions of elevated
activity are advected and stretched, they generate local bursts of vorticity
that seed turbulent motion in previously inactive regions. In this sense, the
activity field acts as a moving, deformable source of turbulence whose spatial
extent and intensity evolve dynamically in time.

The strong coupling between activity heterogeneity and flow structure is
further highlighted in Fig.~\ref{Fig:Fields}(c), which shows the magnitude of the activity
gradient field, $|\nabla \alpha(\mathbf{x},t)|$, at late times. The gradient
field exhibits intermittent, worm-like structures with large amplitudes,
reminiscent of scalar dissipation structures in turbulent flows. These regions
of strong gradients tend to align with zones of intense vorticity and strain,
indicating that sharp activity fronts not only mark transitions between
dynamical regimes, but also actively participate in shaping the local flow.
We refer the reader to Ref.~\cite{fields} which shows the time evolution of these fields 
highlighting the transitions discussed above.

Overall, these results reinforce the interpretation of activity as an
\emph{active scalar}: Its spatial distribution governs where and how turbulent
motion is generated, while the resulting flow feeds back to sharpen gradients,
stretch fronts, and create complex filamentary structures. Even as the system
evolves toward a homogeneous activity state at long times, the transient
dynamics are dominated by this intricate co-evolution of activity and flow.

\subsection{Spectral signatures of heterogeneous activity and transient universality}

We now turn to the spectral consequences of the evolving activity field and its
impact on the structure of active turbulence. As established in
Fig.~\ref{Fig:Contours} and Fig.~\ref{Fig:Fields}, the advected activity field
$\alpha(\mathbf{x},t)$ dynamically partitions the domain into regions that
sustain qualitatively different dynamics. In particular, regions with
$\alpha<\alpha_c\approx -5$ support strongly chaotic and intermittent active
turbulence, while regions with $\alpha>\alpha_c$ remain weakly active and
comparatively quiescent. The spatial extent of the strongly active regions
therefore introduces a time-dependent and physically meaningful length scale
into the problem, which we denote by $R(t)$.

Figure~\ref{Fig:Spectra} shows how this evolving spatial structure is reflected
in the kinetic energy spectra and related statistical measures for a system
initialized with an extreme activity contrast, $\alpha_{\mathrm{in}}=-8$ and $\alpha_{\mathrm{out}}=4$, and with $\rm{Sc} = 1.8$. This
choice is made to clearly expose the mechanism underlying the coexistence of
distinct spectral regimes and the eventual loss of universal behavior;
qualitatively similar trends are also observed for less extreme values (e.g.\
$\alpha_{\mathrm{in}}=-6$ and $\alpha_{\mathrm{out}}=0$), but the non-monotonic evolution of the crossover scale and the
disappearance of the $k^{-3/2}$ regime are most clearly resolved in this case.

Figure~\ref{Fig:Spectra}(a) displays the kinetic energy spectra $E_k$ at four
representative times. At early times, the spectrum exhibits two distinct
scaling regimes. At larger wavenumbers, the spectrum follows the universal
$k^{-3/2}$ scaling previously identified in homogeneous systems with
$\alpha<\alpha_c$, reflecting strongly active turbulent dynamics. At smaller
wavenumbers, however, the spectrum crosses over to a non-universal scaling
associated with weakly active regions. The coexistence of these regimes
indicates that, during this transient phase, the flow simultaneously samples
both strongly and weakly active environments.

The crossover between the two scaling regimes occurs at a time-dependent
wavenumber $k_R(t)\sim R^{-1}(t)$, where $R(t)$ is the instantaneous linear
size of the region satisfying $\alpha<\alpha_c$. This crossover scale is not
imposed externally, but emerges dynamically from the evolving activity field.
Its real-space origin is made explicit in Fig.~\ref{Fig:Spectra}(b), which
shows snapshots of the activity field $\alpha(\mathbf{x},t)$ at the same four
times, with regions where $\alpha<\alpha_c$ marked with a blue colormap. These blue
regions directly define $R(t)$ and coincide with the zones where intense
vorticity and chaotic motion are sustained, as previously seen in
Fig.~\ref{Fig:Fields}.

At early times, the blue region in Fig.~\ref{Fig:Spectra}(b) forms a compact
but growing patch. Length scales smaller than $R(t)$ are therefore dominated by 
fluctuations arising within the strongly active environment and give rise to the
universal $k^{-3/2}$ scaling observed at large wavenumbers in
Fig.~\ref{Fig:Spectra}(a). In contrast, length scales comparable to or larger
than $R(t)$ necessarily sample both strongly and weakly active regions, leading
to a breakdown of universality and a crossover to non-universal scaling. As
time progresses, the snapshots in Fig.~\ref{Fig:Spectra}(b) show that the
spatial extent of the region with $\alpha<\alpha_c$ initially increases due to
advection and stretching of the activity patch. Correspondingly, $R(t)$ grows
and the crossover wavenumber $k_R(t)$ shifts to smaller values, allowing the
universal $k^{-3/2}$ scaling to extend to progressively larger scales in
Fig.~\ref{Fig:Spectra}(a).

At later times, diffusion and mixing progressively smooth out activity
gradients. The blue regions in Fig.~\ref{Fig:Spectra}(b) shrink and eventually
become negligible, indicating that no extended region with $\alpha<\alpha_c$
survives. This erosion of the strongly active core causes the crossover
wavenumber $k_R(t)$ to shift back toward larger values and ultimately
eliminates the $k^{-3/2}$ regime altogether, leaving a single non-universal
spectral scaling once the activity field becomes sufficiently homogeneous. 
While Figs~\ref{Fig:Spectra}(a) and (b) are self-consistent and reflect the 
phenomenological argument developed above, finding a clear quantitative measurement 
of $R(t)$ (and hence $k_R(t)$), and their variation with the Schmidt number, 
is less trivial in such heterogeneous systems and left for future work.

The connection between the spatial evolution of the activity field and the
spectral behavior is further reinforced in the inset of Fig.~\ref{Fig:Spectra}(c), which
shows the probability density functions (PDFs) of the activity field $\alpha$
at the same four times. At early times, the PDFs are close to bimodal,
reflecting the sharp spatial segregation between the strongly active patch and
the weakly active background seen in Fig.~\ref{Fig:Spectra}(b). As time
evolves, the PDFs progressively flatten with vanishing contributions for $\alpha \lesssim \alpha_c (=-5)$, 
while still peaking at the right, indicating enhanced mixing
and homogenization of the activity field. By the latest time, the disappearance
of a tail below $\alpha_c$ confirms that strongly active regions no
longer persist, consistent with the absence of universal scaling in the energy
spectra.

Additional insight into the fluctuations of the activity field is gained by monitoring the 
evolution of the $\alpha-$spectrum $E^\alpha_k = \langle |\alpha_k|^2 \rangle$, 
where $\alpha_k$ are the Fourier modes of the activity field. 
The spectral evolution of the activity field is shown in Fig.~\ref{Fig:Spectra}(c) for the same four times as in
Fig.~\ref{Fig:Spectra}(a) and (b). At early times, the activity spectrum exhibits an approximate $k^{-2}$ scaling over
an intermediate range of wavenumbers. This behavior originates from the presence of sharp,
shock-like interfaces separating strongly and weakly active regions, as visible in the corresponding
activity snapshots in Fig.~3(b). Such discontinuity-dominated structures naturally give rise to
$k^{-2}$ spectra~\cite{MuruganPRR}.

At later times, advection and diffusion progressively smooth these sharp activity gradients, and
the spectrum steepens toward a $k^{-4}$ decay, consistent with an increasingly smooth activity
field. Further, an intermediate $k^{-1}$ regime is observed. While this scaling is
reminiscent of the Batchelor regime for passively advected scalars~\cite{Batchelor1959}, we emphasize that activity here
is not passive and directly feeds back onto the flow; the resemblance is therefore purely
qualitative at this stage and any equivalence between the two is left for future work.

This interpretation is further supported by the inset of Fig.~\ref{Fig:Spectra}(c), which shows the probability
density functions (PDFs) of the activity field at the same times. At early times, the PDFs are
near-bimodal, reflecting the coexistence of strongly active ($\alpha<\alpha_c$) and weakly active
regions. As time progresses, the distribution smoothens and the low-$\alpha$ tail progressively
disappears. At late times, the PDFs show negligible probability for
$\alpha\lesssim\alpha_c\simeq -5$, indicating the erosion of strongly active patches. Consistently,
the universal $k^{-3/2}$ scaling in the kinetic energy spectrum [Fig.~\ref{Fig:Spectra}(a)] vanishes at these times,
confirming that the loss of strongly active regions directly eliminates the universal turbulent
regime.

Taken together, Fig.~\ref{Fig:Spectra} demonstrates that the emergence,
evolution, and eventual disappearance of universal $k^{-3/2}$ scaling in
heterogeneous active turbulence can be traced directly to the spatiotemporal
evolution of the activity field. The key control parameter is not the globally
averaged activity (which remains constant), but the instantaneous size $R(t)$ of regions satisfying
$\alpha<\alpha_c$. Universality is therefore local in space and transient in
time: It appears only on scales smaller than $R(t)$ and persists only as long
as strongly active patches survive. As advection, diffusion, and mixing reshape
the activity field, the window of universal behavior shifts dynamically across
scales and ultimately closes once the activity becomes homogeneous. Naturally, this window becomes shorter with a 
decreasing $\rm{Sc}$ number which accelerates activity homogenization.

\section{Conclusions and discussions}
\label{sec:Conclusions}

In this work, we have presented a combined review and theoretical study of
turbulence-like states in dense active fluids, with particular emphasis on
bacterial suspensions and related systems described by
Toner--Tu--Swift--Hohenberg-type hydrodynamics. Building on earlier
experimental and theoretical developments, we revisited the analogy between
active turbulence and classical inertial turbulence, clarified its domain of
validity, and highlighted the distinct nonequilibrium mechanisms that underlie
chaotic flow generation in active matter.

A central message of this study is that active turbulence does not constitute a
single universal state, but instead spans multiple dynamical regimes even simultaneously, controlled
by activity, confinement, and nonlinear interactions. While homogeneous active
systems can exhibit a transition to a strongly chaotic regime with universal
spectral behavior and intermittency, real active systems are often spatially
heterogeneous. Motivated by this observation, we focused on the role of
spatially varying activity and introduced a minimal framework in which activity
is treated as a dynamically evolving field advected by the flow it generates.

Our numerical results demonstrate that an advected activity field gives rise to
sharp activity fronts, transient confinement of turbulent motion, and complex
interfacial morphologies. These fronts act as dynamical boundaries
that partition the domain into regions with qualitatively distinct dynamics. As
a consequence, energy spectra can exhibit coexisting scaling regimes associated
with strongly and weakly active regions, with a crossover scale set by the
instantaneous spatial extent of the strongly active patches. As activity is
mixed and homogenized by advection and diffusion, this crossover shifts in time
and the universal regime may ultimately disappear. Universality in active
turbulence is therefore inherently local in space and transient in time in
heterogeneous systems.  The emergence of strongly heterogeneous and dynamically
evolving flow structures in the presence of advected activity suggests
additional physical consequences beyond spectral statistics. In classical
inertial turbulence, coherent flow structures are known to play a central role
in collision rates, coalescence, and transport processes~\cite{Picardo2019,Picardo2020}. Our results raise the
possibility that analogous structure-mediated interactions may arise in
biological active flows, where activity heterogeneity could indirectly regulate
encounters, mixing, and collective behavior.

An important direction for future work concerns the mathematical structure and
higher-dimensional generalizations of the hydrodynamic equations underlying
active turbulence. Recent analytical studies have begun to rigorously
characterize the Toner--Tu and Toner--Tu--Swift--Hohenberg equations,
addressing questions of well-posedness, global attractors, and bounds on
solutions in two dimensions \cite{Pandit2025,Gibbon2023,Boutros2026}. Such
results provide a valuable theoretical foundation for the phenomenological and
numerical observations reported here, particularly in regimes where strong
activity and nonlinear advection lead to highly chaotic dynamics. Extending
these mathematical analyses to settings with spatially heterogeneous and
dynamically evolving activity fields remains an open challenge.

Equally important is the extension of active turbulence studies beyond
quasi-two-dimensional geometries. While most experimental and theoretical work,
including the present study, has focused on effectively two-dimensional
systems, recent investigations have explored active turbulence in fully
three-dimensional fluids~\cite{alert2022active,perlekar2026}. Understanding how activity heterogeneity, front
dynamics, and transient universality manifest in three dimensions is a natural
and necessary next step. Progress along these lines would help bridge minimal
hydrodynamic models with more realistic biological and synthetic active systems
and clarify the extent to which the concepts of confinement, crossover scales,
and local universality identified here persist in higher dimensions.

In summary, treating activity as a dynamical, spatially structured field
reveals new mechanisms by which turbulence-like states in active matter are
organized, sustained, and ultimately destroyed. By explicitly accounting for
the interplay between advection, diffusion, and activity-dependent forcing,
this work provides a framework for understanding active turbulence in more
realistic, heterogeneous environments, and highlights the central role of
activity fronts and interfaces in shaping non-equilibrium chaotic flows.

\begin{acknowledgements}
	SM and SSR thank the Indo–French Centre for Applied
Mathematics (IFCAM) for financial support. The
simulations were performed on the ICTS clusters \emph{Tetris} and
\emph{Contra}. SM acknowledges the IITK Initiation Grant project IITK/ME/2024316 and the Govt. of India grant ANRF/ECRG/2024/002467/ENS. SM would like to thank ICTS-TIFR for their hospitality and support. SS and SSR acknowledge support of the Department of Atomic Energy, Government of India, under project no. RTI4019.
\end{acknowledgements}

\bibliography{references}

\end{document}